\documentclass[12pt]{article}
\usepackage{epsfig,url}

\newcommand\pubnumber{SLAC--PUB--10185}
\newcommand\pubdate{November, 2003}
\newcommand\hepnumber{hep-ph/0310039}

\def\SLAC{Stanford Linear Accelerator Center, Stanford University\\
    2575 Sand Hill Road, Menlo Park, CA 94025  USA}
\def\doeack{\footnote{Work supported by the Department of Energy,
                     contract DE--AC03--76SF00515.}}
\def\maxack{\footnote{Work supported by the National Science 
Foundation under grant PHY-9513717.}}

\def\Title#1{\begin{center} {\Large #1 } \end{center}}
\def\Author#1{\begin{center}{ \sc #1} \end{center}}
\def\Address#1{\begin{center}{ \it #1} \end{center}}
\def\andauth{\begin{center}{and} \end{center}}
\def\submit#1{\begin{center}Submitted to {\sl #1} \end{center}}
\newcommand\pubblock{\rightline{\begin{tabular}{r} \pubnumber\\
         \pubdate \\ \hepnumber \end{tabular}}}
\newenvironment{Abstract}{\begin{quotation} \begin{center}
                       ABSTRACT
     \end{center}\bigskip  }{\end{quotation}}

\def\submit#1{\begin{center}Submitted to {\sl #1} \end{center}}
\def\Acknowledgements{\bigskip  \bigskip \begin{center} \begin{large}
             \bf ACKNOWLEDGEMENTS \end{large}\end{center}}

\def\beq{\begin{equation}}
\def\eeq#1{\label{#1}\end{equation}}
\def\eeqn{\end{equation}}

\newenvironment{Eqnarray}%
   {\arraycolsep 0.14em\begin{eqnarray}}{\end{eqnarray}}
\def\beqa{\begin{Eqnarray}}
\def\eeqa#1{\label{#1}\end{Eqnarray}}
\def\eeqan{\end{Eqnarray}}
\def\CR{\nonumber \\ }

\def\leqn#1{(\ref{#1})}

\let\bar=\overbar

\def\VEV#1{\left\langle{ #1} \right\rangle}

\def\lsim{\mathrel{\raise.3ex\hbox{$<$\kern-.75em\lower1ex\hbox{$\sim$}}}}
\def\gsim{\mathrel{\raise.3ex\hbox{$>$\kern-.75em\lower1ex\hbox{$\sim$}}}}

\def\tr{{\mbox{\rm tr}}}
\def\half{\frac{1}{2}}

\def\tthird{\frac{2}{3}}

\def\del{\partial}
\def\Dslash{\not{\hbox{\kern-4pt $D$}}}
\def\dslash{\not{\hbox{\kern-2pt $\del$}}}

\def\mz{m_Z}

\def\mw{m_W}
\def\mt{m_t}

\def\msb{{\bar{\ssstyle M \kern -1pt S}}}

\makeatletter
\def\section{\@startsection{section}{0}{\z@}{5.5ex plus .5ex minus
 1.5ex}{2.3ex plus .2ex}{\large\bf}}
\def\subsection{\@startsection{subsection}{1}{\z@}{3.5ex plus .5ex minus
 1.5ex}{1.3ex plus .2ex}{\normalsize\bf}}
\def\subsubsection{\@startsection{subsubsection}{2}{\z@}{-3.5ex plus
-1ex minus  -.2ex}{2.3ex plus .2ex}{\normalsize\sl}}

\renewcommand{\@makecaption}[2]{%
   \vskip 10pt
   \setbox\@tempboxa\hbox{\small #1: #2}
   \ifdim \wd\@tempboxa >\hsize     
       \small #1: #2\par          
     \else                        
       \hbox to\hsize{\hfil\box\@tempboxa\hfil}
   \fi}

 \def\citenum#1{{\def\@cite##1##2{##1}\cite{#1}}}
 
\newcount\@tempcntc
\def\@citex[#1]#2{\if@filesw\immediate\write\@auxout{\string\citation{#2}}\fi
  \@tempcnta\z@\@tempcntb\m@ne\def\@citea{}\@cite{\@for\@citeb:=#2\do
    {\@ifundefined
       {b@\@citeb}{\@citeo\@tempcntb\m@ne\@citea\def\@citea{,}{\bf ?}\@warning
       {Citation `\@citeb' on page \thepage \space undefined}}%
    {\setbox\z@\hbox{\global\@tempcntc0\csname b@\@citeb\endcsname\relax}%
     \ifnum\@tempcntc=\z@ \@citeo\@tempcntb\m@ne
       \@citea\def\@citea{,}\hbox{\csname b@\@citeb\endcsname}%
     \else
      \advance\@tempcntb\@ne
      \ifnum\@tempcntb=\@tempcntc
      \else\advance\@tempcntb\m@ne\@citeo
      \@tempcnta\@tempcntc\@tempcntb\@tempcntc\fi\fi}}\@citeo}{#1}}
\def\@citeo{\ifnum\@tempcnta>\@tempcntb\else\@citea\def\@citea{,}%
  \ifnum\@tempcnta=\@tempcntb\the\@tempcnta\else
  {\advance\@tempcnta\@ne\ifnum\@tempcnta=\@tempcntb \else\def\@citea{--}\fi
    \advance\@tempcnta\m@ne\the\@tempcnta\@citea\the\@tempcntb}\fi\fi}
\makeatother

\def\orderof#1{{\mathcal O} \left(#1 \right)}
\def\Br{{\rm Br}}

\def\Cornell{Newman Laboratory for Elementary Particle Physics, \\
Cornell University, 
Ithaca, NY~14853 USA}

\def\stacksymbols #1#2#3#4{\def\theguybelow{#2}
    \def\vp{\lower#3pt}
    \def\sp{\baselineskip0pt\lineskip#4pt}
    \mathrel{\mathpalette\intermediary#1}}

\def\intermediary#1#2{\vp\vbox{\sp
     \everycr={}\tabskip0pt
     \halign{$\mathsurround0pt#1\hfil##\hfil$\crcr#2\crcr
              \theguybelow\crcr}}}

\def\lapproxeq{\stacksymbols{<}{\sim}{2.5}{.2}}

\begin{document}
\begin{titlepage}
\pubblock

\vfill
\Title{Top Quarks and Electroweak Symmetry Breaking in Little Higgs Models}
\vfill
\Author{Maxim Perelstein\maxack}
\Address{\Cornell}
\andauth
\Author{Michael E. Peskin and Aaron Pierce\doeack}
\Address{\SLAC}
\vfill
\begin{Abstract}
`Little Higgs' models, in which the Higgs particle
 arises as a pseudo-Goldstone boson,
have a natural mechanism of electroweak symmetry breaking associated with 
the large value of the top quark Yukawa coupling.  The mechanism typically 
involves a new heavy $SU(2)_{L}$ singlet top quark, $T$.  We discuss the 
relationship of the Higgs boson and the two top quarks.  We suggest 
experimental tests of the Little Higgs mechanism of electroweak symmetry 
breaking using the production and decay of the $T$ at the Large Hadron
Collider.
\end{Abstract}
\vfill
\submit{Physical Review {\bf D}}
\vfill
\end{titlepage}
\tableofcontents
\newpage
\def\thefootnote{\fnsymbol{footnote}}
\setcounter{footnote}{0}

\section{Introduction}

The most pressing question in elementary particle physics today is that of 
identifying the mechanism responsible for the spontaneous breaking of the 
$SU(2)\times U(1)$ symmetry of the weak interactions.  For many years, the 
list of candidate answers to this question was static.  The leading 
alternatives were supersymmetry and new strong interactions at the 
TeV scale.  Recently, the list has expanded to include several 
new mechanisms, including candidates gleaned from analyses of models 
with extra dimensions.  Whatever the mechanism of electroweak symmetry 
breaking, we expect it to be associated with the TeV scale, which we 
will explore soon at the Large Hadron Collider (LHC).  It is important, 
then, to clarify the implications of these new mechanisms and 
the observable processes by which they might be tested.

One of the most appealing of the newly proposed approaches to 
electroweak symmetry breaking is that of the
`Little Higgs'~\cite{LHorig,LHorig2,LH1,LH2,Schmaltz,Wacker}.  This model
revives the idea that the Higgs particle is a pseudo-Goldstone
boson~\cite{Georgi,GeorgiKaplan,KGS}, adding to it a number of insights from 
the study of extra dimensions, supersymmetry and other 
weak-coupling Higgs theories.  
Proponents of the Little Higgs argue that the large top Yukawa coupling 
can generate the instability of the Higgs potential to electroweak
symmetry breaking. The construction links the observed heaviness of 
the top quark to electroweak symmetry breaking in a manner different 
from that in supersymmetry~\cite{SUSYt1,SUSYt2,SUSYt3,SUSYt4} or 
topcolor~\cite{topcolor} models, through a mechanism that is direct and 
appealing. In this paper, we will discuss the relation between this 
mechanism and the properties of the top quark and its partners.

Little Higgs models typically contain a large multiplet of pseudo-Goldstone
bosons, including the Higgs doublet of the Standard Model. 
While many of the Goldstone bosons in this multiplet receive masses at the TeV 
scale, the models are constructed so that the Higgs boson mass is 
protected from quadratic divergences at the one-loop level. The dominant 
contributions to the Higgs boson mass parameter are only logarithmically 
sensitive to the physics at the cutoff and are therefore calculable. 
The (mass)$^2$ parameter generated by gauge interactions in perturbation 
theory is positive.  However, the couplings of the Higgs boson
to the top quark and to a new heavy vector-like quark 
can overcome the positive contribution, and produce total (mass)$^2$ 
parameter for the Higgs doublet that is negative.  Therefore, like 
supersymmetry and 
topcolor, the explanation of the negative (mass)$^2$ of the Higgs boson 
is tied to its couplings to the top sector.  However, there is an  
advantage that Little Higgs models have with respect to supersymmetry.  
In supersymmetry, the calculation of electroweak symmetry breaking combines 
the contribution from the top sector with the independent parameters $\mu$ and
B$\mu$, whereas in the Little Higgs model the top contribution stands on its
own.  

In the Little Higgs model, the couplings of the Higgs to the Standard Model 
top quark, $t$, and the new heavy top quark, $T$, form an independent 
sector that is relatively isolated from the rest of the Higgs 
dynamics.  This allows us to make statements about the dynamics of the 
$T$ that are general in models making use of the 
Little Higgs mechanism.  Tests of these statements test the underlying 
mechanism of electroweak symmetry breaking.

In this paper, we will consider only models with one new heavy top and 
one pseudo-Goldstone boson Higgs doublet.  Our conclusions are general 
with these assumptions.  More complicated 
top sectors and models with multiple Higgs doublets have been 
proposed~\cite{Nelson,Sp6,SK,Jay,Sp6Precision,Skiba:2003yf,Spencer}.

The plan of this paper is as follows:   In Section 2, we will review the 
mechanism of electroweak symmetry breaking in Little Higgs models.  
From this discussion, we will obtain a relation among the parameters of the 
Lagrangian that couples the Higgs to the top quarks.  The rest of our 
discussion will be concerned with methods of testing this relation.   
In Section 3, we will discuss the parameters of a simple version of the 
Little Higgs model and the constraints on these parameters from precision
electroweak measurements. This discussion will build on the work of 
\cite{Csaki,HPR,Csaki2}. The goal of this section will be to determine the
acceptable values of the $T$ mass.
In Section 4, we will discuss the phenomenology of the $T$.  We will
argue that, though the $T$ is to first order a weak interaction singlet, 
it decays significantly to  $W^+ b$ and $Z^0 t$.  These modes provide 
important signatures for $T$ production at the LHC.  We will argue 
that the measurement of the total width of the $T$ and of its production 
cross section tests the relation highlighted in Section 2.  Section 5 
presents some conclusions.

Some other aspects of the Little Higgs model collider phenomenology 
have been discussed in \cite{AaronMaxim,Han,Han2}.

\section{Electroweak symmetry breaking in Little Higgs models}

As we have already noted, electroweak symmetry breaking in Little Higgs models
can result from coupling the Little Higgs multiplet to an isolated sector
containing the top quark and another heavy quark.  In this section, we will 
review this mechanism as it was presented in~\cite{LH1,LH2}.  We will not be 
concerned with the entire computation of the Higgs potential, only with the 
generation of one term in which the Higgs (mass)$^2$ is negative.  We will 
see that the mechanism of ~\cite{LH1,LH2} is a simple and attractive way to 
meet that goal.
The mechanism involves an additional heavy charge-$\tthird$ quark.  
The idea that a heavy singlet quark mixing with the top quark as a part of
electroweak symmetry breaking was originally introduced as the `topcolor
seesaw' of Dobrescu and Hill~\cite{DHill}.
In the 
following, we will use the letter $u$ or $U$ to denote weak eigenstates 
and $t$ or $T$ to denote mass eigenstates.  Then, the third-generation 
weak doublet will be $(u,b)_L$, the new left-handed weak singlet will 
be $U_L$, and the two right-handed weak singlets of the model will be 
$u_R$, $U_R$.   We will identify the $t$ and $T$ states momentarily. 

A key feature of the Little Higgs construction is the presence of global 
symmetries which protect the Higgs
boson mass against quadratically divergent radiative corrections at 
one-loop.  The Higgs boson couplings to quarks should preserve
this feature.  As a demonstration of how this could work, 
we introduce an $SU(3)$ global symmetry.  
Let $V$ be an $SU(3)$ unitary matrix, depending on Goldstone boson 
fields $\pi^a$ as
\beq
             V = \exp [2 i \pi^a t^a/f ] \  ,
\eeq{theV}
where $f$ is a ``pion decay constant'' with the dimensions of mass 
and $t^a$ is an $SU(3)$ generator, normalized 
to $\tr[t^a t^b] = \half \delta^{ab}$. We will identify the 
Higgs doublet $H \equiv (h +i \pi_{3}, \, -\sqrt{2} \pi^{-})^{T}$ 
with the $SU(2)$ doublet 
components of the Goldstone boson matrix $\Pi \equiv \pi^a t^a$:
\beq
           2 i \Pi = \frac{1}{\sqrt{2}} 
\pmatrix{   \Phi  &   H \cr   -H^\dagger   &\phi\cr   }.
\eeq{Pidecomp} 
$\Phi$ and $\phi$ are other members of the Goldstone multiplet
that we need not concern ourselves with at this point.
Let $\chi_L$ be the `royal' $SU(3)$ triplet $(u,b,U)_L$~\cite{Jarry}.  
These fields can be coupled by writing~\cite{LH1,LH2}
\beq
  \mathcal{L} = 
   - \lambda_1 f\, \bar{u}_R V_{3i} {\chi_L}_i  - 
	\lambda_2 f\, \bar{U}_R U_L + {\rm h.c.}\ 
\eeq{thedelta}
The first term of this Lagrangian has an $SU(3)$ global symmetry
\beq
 V_{3i} \to  V_{3j} \Lambda^\dagger_{ji},  \qquad   \chi_L \to \Lambda 
\chi_L \ .
\eeq{theSUthree}
This symmetry is spontaneously broken.
To the extent that this $SU(3)$ is an exact symmetry
of the Lagrangian, the Goldstone boson fields, $\pi^a$,
must remain massless.  The second term in \leqn{thedelta} explicitly breaks the
$SU(3)$ symmetry to $SU(2)$ and specifically breaks the symmetries 
responsible for keeping $H$ and $H^\dagger$ in \leqn{Pidecomp} massless.  
However, the Higgs boson field does not enter this term directly.
This means that $H$ can 
obtain mass only from loop diagrams, and only at a level at which the couplings
$\lambda_1$ and $\lambda_2$ both enter.  In \cite{LH1}, it is shown that this
restriction prohibits the appearance of one-loop quadratic divergences in 
the Higgs boson mass. The one-loop radiative contribution to the Higgs 
(mass)$^2$ is only logarithmically divergent, and can thus be reliably 
estimated. This contribution turns out to be negative~\cite{LH1}, giving an 
explicit mechanism of electroweak symmetry breaking. 

\begin{figure}[t]
\begin{center}
\epsfig{file=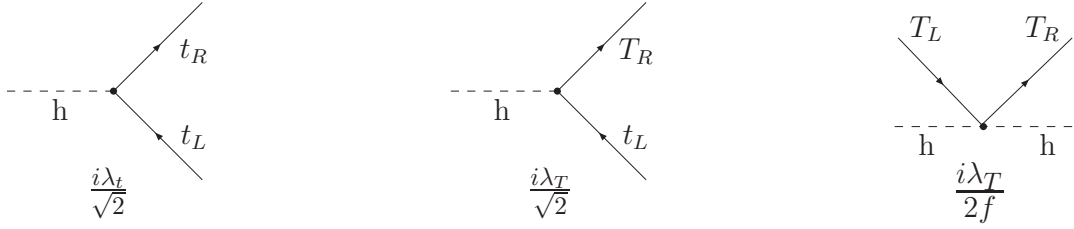,width=\columnwidth}
\caption{Feynman rules for couplings between the 
Higgs boson and the top quarks in the symmetric vacuum 
($\langle h \rangle = 0$).  We have only shown those vertices relevant
to the calculation of the one-loop quadratic divergences from the top sector.
There are additional vertices,
generated by terms of higher order in the expansion of $V$, involving
three or more Higgs bosons.}
\label{fig:zerorules}
\end{center}
\end{figure}

Let us review both aspects of the calculation.
We expand about the symmetric point,
$\VEV{h} = 0$.  At this point,  $u_L$ remains massless, while $U_L$ combines
with one linear combination of $u_R$ and $U_R$ to obtain a mass.  The 
mass eigenstates are then
\beqa
  t_L = u_L,       &\qquad&   t_R =  {\lambda_2 u_R-  \lambda_1 U_R\over
                      \sqrt{\lambda_1^2 + \lambda_2^2}}, \CR
   T_L = U_L,      & \qquad&  T_R = {\lambda_1 u_R + \lambda_2 U_R\over
                                   \sqrt{\lambda_1^2 + \lambda_2^2}},
\eeqa{teigenstates}
with $m_{t}$ massless at this level and 
\beq
m_{T}= \sqrt{\lambda_{1}^{2} +\lambda_{2}^{2}} \, f.
\eeq{heavytopmass}
The Feynman rules for couplings between the Higgs boson and the top quarks 
in the symmetric vacuum are given in 
Fig.~\ref{fig:zerorules}.  We only show rules involving one or two Higgs 
bosons, which are relevant to the calculation of the one-loop quadratic 
divergence. 
The couplings of the Higgs boson to $t_L \bar t_R$ and to 
$t_L \bar T_R$ are related to the parameters appearing in the 
Lagrangian~\leqn{thedelta} via
\beq
     \lambda_t = {\lambda_1\lambda_2 \over   \sqrt{\lambda_1^2 + \lambda_2^2}}
            \ ,  \qquad 
     \lambda_T = {\lambda_1^2 \over   \sqrt{\lambda_1^2 + \lambda_2^2}}\ .
\eeq{tcouplings}

\begin{figure}[t]
\begin{center}
\epsfig{file=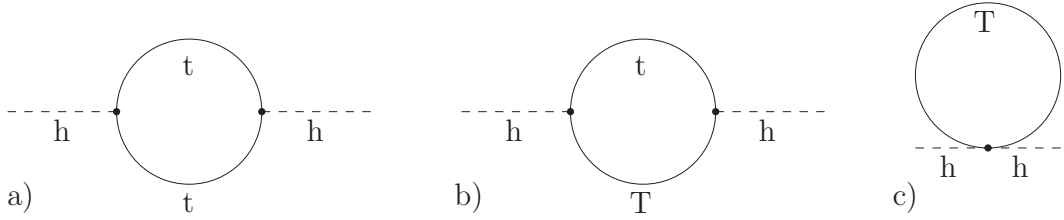,width=\columnwidth}
\caption{One-loop contributions to the Higgs boson (mass)$^{2}$ in the Little
Higgs model.}
\label{fig:oneloop}
\end{center}
\end{figure}

The one-loop contribution to the Higgs boson (mass)$^2$ comes from the 
three diagrams in Fig.~\ref{fig:oneloop}. The values of the diagrams are
\beqa
     \mbox{a)} &=&  -6 \lambda_t^2 \int {d^4k\over (2\pi)^4}  {1\over k^2},\CR
     \mbox{b)} &=&  -6 \lambda_T^2 \int {d^4k\over (2\pi)^4}
                                           {1\over k^2- m_T^2},\CR \CR
     \mbox{c)} &=& +6 {\lambda_T\over f}
            \int {d^4k\over (2\pi)^4}  {m_T\over k^2 - m_T^2}.
\eeqa{tloops} 

The quadratic divergences neatly cancel. The top sector contribution to the 
Higgs (mass)$^{2}$ is then given by 
\beq
   \Delta m_h^2 =  - 3 { \lambda_1^2 \lambda_2^2 f^2\over 8\pi^2 }  
                   \log {\Lambda^2\over m_T^2} = 
            - 3 { \lambda_t^2 m_{T}^{2} \over 8\pi^2 }  
                   \log {\Lambda^2\over m_T^2} \ ,
\eeq{higgsmassanswser}
where $\Lambda \sim 4\pi f$ is the strong interaction scale of the theory 
that gives rise to the Goldstone bosons.   In Little Higgs models, $f$ is 
typically taken to be of order 1 TeV (corresponding to $\Lambda \sim 10$ TeV) 
to avoid fine tuning of the Higgs mass.   As long as  $m_T$ is parametrically 
lower than $\Lambda$, the negative contribution to $m_h^2$ in 
Equation~\leqn{higgsmassanswser} could be the dominant one and thus would 
provide the explanation for why electroweak symmetry is broken. There are 
incalculable (quadratically divergent) two-loop contributions to $m_{h}^{2}$, 
which are the same order in $\lambda_{1} \lambda_{2}$, but these are not 
logarithmically enhanced, and so are sub-dominant.  The situation is that 
typically found in chiral perturbation theory.

The cancellation of quadratic divergences in Equation~\leqn{tloops} depends 
on the 
relation of Equation~(\ref{heavytopmass}), which can be rewritten as
\beq
     \frac{m_T}{f} =    { \lambda_t^2  + \lambda_T^2\over  \lambda_T}\ .
\eeq{Tmasstwo}
The relation \leqn{Tmasstwo} is a very interesting one.  
All of the four parameters
in this equation are in principle measurable.  The top quark Yukawa coupling
is known.  The decay constant $f$ can be determined by measuring the 
properties of the heavy vector bosons in the 
Little Higgs theory \cite{AaronMaxim}.
 The mass and couplings
of the heavy top quark will be measured when this quark is observed, perhaps
at the LHC.   If the relation \leqn{Tmasstwo} is shown to be valid, 
that will be 
strong evidence for the picture of electroweak symmetry breaking given by 
the Little Higgs model.

\section{How heavy is the heavy top quark?}
\label{PEW}

If we are to study the heavy quark $T$, we should have some idea of what its
mass is.  The Little Higgs theory does not place a firm upper bound on the mass
of the $T$.  However, if the mass of the $T$ is greater than about 2 TeV, 
the cancellation
shown in~\leqn{tloops} requires some tuning to give an answer for $m_h$ 
below 220 GeV, the range for the Higgs boson mass preferred by precision 
electroweak measurements~\cite{EWWG}.  For this reason, the 
authors of~\cite{LH1} 
suggested that the mass of the $T$ should be less than 2 TeV.  

On the other hand, the consistency of the Little Higgs model with precision 
electroweak data can place a lower limit on the mass of the $T$.   
The precision
electroweak corrections from the model of \cite{LH2} have been studied in 
some detail in \cite{Csaki} and \cite{HPR}.  These authors have found very 
strong
bounds that imply  $m_T > 8$--10 TeV.  The corrections from $T$ loops were
computed in \cite{Csaki,HPR}, but these turned out to be relatively 
unimportant terms
 of order $\alpha m_t^4/m_T^2$.  The large 
effects are the direct tree-level modifications of the precision electroweak 
predictions due to the new heavy vector bosons in the Little Higgs model.
Consideration of their effects gives a lower bound on $f$.  To find a 
limit on the mass of the $T$, we can use such a bound in conjunction with 
the inequality
\beq
            \frac{m_T}{f}  \geq 2 \lambda_{t} \approx 2,
\eeq{mTlower}
which is obtained by minimizing \leqn{Tmasstwo} with respect to $\lambda_T$.

There is a specific reason that the model of \cite{LH2} leads to a very strong
bound on $f$.  In this model, all quadratically divergent contributions to 
$m_h^2$
due to $W$ and $Z$  boson loops are canceled by contributions from heavy
gauge bosons.  To achieve this, the authors of \cite{LH2}  make use of a gauge
group $SU(2)\times SU(2) \times U(1) \times U(1)$.  This leads to a multiplet
of heavy $SU(2)$ gauge bosons and a heavy $U(1)$ gauge boson.  The 
heavy $U(1)$ boson is actually not very heavy, 
\beq
         m \sim    \frac{2}{\sqrt{5}} {g' f} \ \approx 0.3\,f,
\eeq{lowmassZ}
and this leads to large electroweak corrections,  and also to 
problems with the direct
observational bounds on $Z'$ bosons from the Tevatron \cite{Csaki,HPR}.

\subsection{An $SU(2)\times SU(2) \times U(1)$ model}

There are various ways to ameliorate this problem~\cite{Csaki2}, but the 
most direct is to gauge a smaller group.  
It has been suggested~\cite{AHWsugg} that one 
might gauge only $SU(2)\times SU(2)\times U(1)$, canceling the quadratic 
divergences proportional to $g^2/4\pi$ but allowing quadratically divergent
terms proportional to $g^{\prime 2}/4\pi$.  Since $g^{\prime}$  is small, 
the latter are not unreasonably 
large if the cutoff or strong interaction scale $\Lambda$ of the Little 
Higgs model is about 10 TeV.
In the remainder of this section, we will adopt this approach to find a more 
conservative lower bound on $f$ and on the $T$ quark mass.

The success of this approach depends on the exact choice of the 
symmetry-breaking pattern that produces the pseudo-Goldstone 
bosons of the Little Higgs models.  Given the importance of the global 
$SU(3)$ symmetry for the top couplings, one might first study a model in 
which a global $SU(3)\times SU(3)$ symmetry is spontaneously broken 
to $SU(3)$.  The multiplet of 
 Goldstone bosons fill an adjoint representation of $SU(3)$.  The Higgs field 
can be identified as an $SU(2)$ doublet within this structure, 
\beq
              2i \pi^a t^a = \frac{1}{\sqrt{2}} 
                     \pmatrix{ 0 & 0 &   h+i \pi^3\cr 
                     0 & 0 & - \sqrt{2} \pi^-\cr
                    -(h-i\pi^3) & \sqrt{2} \pi^+ & 0 \cr}\  .
\eeq{HinSUthree}
Exponentiating and taking the vacuum expectation value 
$\langle h \rangle = v=246$ GeV, we find the $SU(3)$ nonlinear sigma model 
field
\beq
    V = e^{2i\pi^at^a/f}  =
            \pmatrix{
        \cos \frac{v}{\sqrt{2} f} & 0 & \sin \frac{v}{\sqrt{2}{f}}  \cr 
        0 & 1 & 0 \cr 
-\sin \frac{v}{\sqrt{2}{f}} & 0 & \cos \frac{v}{\sqrt{2} f} \cr},
\eeq{theVforthree}
The kinetic Lagrangian for this field is
\beq 
    \mathcal{L}   =    {f^2\over 2} \tr\left[  D_\mu V^\dagger D^\mu V \right] 
\eeq{theLforthree}
with 
\beq 
      D_\mu V =  \del_\mu V - i g_L A_{L \mu}^a T^a V + i g_R A_{R\mu}^a 
                   V T^a   - i g' B_\mu [Q, V]. 
\eeq{theDforthree}
Here, $T^a=\,$diag$\,(\tau^a, 0)$, where $\tau^a$ is an $SU(2)$ generator, 
and $Q$ is a matrix of $U(1)$ charges with $(-\half, \half, 0)$
on the diagonal. Using these formulae, it is straightforward to work out the 
vector boson masses.  To leading order in $v/f$, the 
heavy triplet of $W$'s have masses given by:
\beq
m_{W_{H}}^{2}=\frac{g_{L}^{2}+ g_{R}^{2}}{2} f^{2}.
\eeq{Wheavythree}
The masses of the usual $W$ and $Z$ turn out to be related by 
\beq
 \mw^2/\mz^2 = \cos^2\theta \left( 1 + {1\over 8} {v^2\over f^2} \right) \ ,
\eeq{WZinthree}
where $\cos^{2} \theta$ is the weak mixing angle defined in terms of the 
underlying gauge coupling constants.
This gives an unacceptable violation of the known $W$/$Z$ mass relation
if  $f < 3$ TeV.  The problem stems from the fact that this model does not 
respect custodial $SU(2)$ at the level of $v^2/f^2$ corrections, as was 
pointed out already in the original papers of Georgi and Kaplan on
 pseudo-Goldstone models for the Higgs  boson~\cite{GKSUtwo}.

The symmetry breaking pattern $SU(5)/SO(5)$, which is the basis of \cite{LH2}, 
is much more promising from this point of view.  In the old approach of Kaplan
and Georgi, this model preserves custodial $SU(2)$.  When we gauge 
$SU(2)\times SU(2)\times U(1)$  as in the Little Higgs models, custodial 
$SU(2)$ is explicitly broken, but it is possible to check that the
custodial $SU(2)$-violating corrections to the vector boson mass relation
\leqn{WZinthree} appear for the first time in order $v^6/f^6$.  With this 
problem and that of the (too--light) heavy $U(1)$ boson removed, there is 
no further reason for a major difficulty with the precision electroweak data.

In the remainder of this section, we give some details of a more thorough
analysis of this question.\footnote{A similar analysis has been 
presented in~\cite{Csaki2}.} To begin, the Higgs doublet must be fit into the 
multiplet of Goldstone bosons of $SU(5)$ spontaneously broken to $SO(5)$.
To do this, we write
\beq
 2i \pi^a t^a =  {1\over \sqrt{2}} \pmatrix{ 0 & 0 &  ( h+i \pi^3 )& 0 & 0 \cr
                   0 & 0 & \sqrt{2} \pi^-  & 0 & 0 \cr
 -( h-i\pi^3 )& - \sqrt{2} \pi^+ & 0  &  -( h+i\pi^3 )&  -\sqrt{2} \pi^- \cr
                0 & 0 &  ( h-i \pi^3 )& 0 & 0 \cr
                0 & 0  & \sqrt{2} \pi^+  & 0 & 0 \cr}\  ,
\eeq{HinSUfive}
where we only show the degrees of freedom corresponding to the Higgs 
doublet.\footnote{The remaining physical (uneaten) degrees of freedom in the
Goldstone boson multiplet form a triplet under $SU(2)_L$, and obtain a mass
at the TeV scale via radiative corrections.}

It is convenient to take the vacuum configuration to be 
\beq
        V_{0} = \pmatrix{ 0 & 0 & 0& 1 & 0 \cr
                                          0 & 0 & 0& 0 & 1 \cr 
                                           0 & 0 & 1& 0 & 0 \cr 
                                            1& 0 & 0& 0 & 0 \cr
                                           0 & 1 & 0& 0& 0 \cr} \ .
\eeq{Visone}
Then, exponentiating the action of $\pi^at^a$, we find the $SU(5)$
nonlinear sigma model field
\beq
    V = e^{2i\pi^at^a/f} V_{0}  = \pmatrix{
       -\half(1-c) & 0 & {s/\sqrt{2}}  & \half(1+c) & 0 \cr 
          0  & 0 &  0 & 0   & 1\cr 
          -{s/ \sqrt{2}} & 0 & c  & -{s/ \sqrt{2}} & 0 \cr
         \half(1+c) & 0 & {s/ \sqrt{2}}  & -\half(1-c) & 0 \cr 
                                           0 & 1 & 0& 0& 0 \cr} \ ,
\eeq{theVforfive}
with 
\beq
    c = \cos{v\over f}  \ , \qquad   s = \sin {v\over f} \ .
\eeq{thesincos}

We now gauge the $SU(2)$ acting on the first two rows and columns of $V$ 
with gauge coupling $g_L$, the $SU(2)$ acting on the last two rows and 
columns of $V$ with gauge coupling $g_R$, and the unbroken $U(1)$ with
coupling $g'$. The normalization of the gauged $U(1)$ generator is chosen 
to ensure the correct value of the Higgs boson hypercharge. 
The kinetic Lagrangian for the $V$ field is
\beq
    \mathcal{L}  = {f^2\over 4} \tr \left[  D_\mu V^\dagger  D^\mu V\right] \ .
\eeq{Lforfive}
Here the covariant derivative of the $V$ field is given by
\beqa
D_{\mu} V = \partial_{\mu} V &-& i\sum_{j=L,R} g_{j} A^{a}_{j}
(Q^{a}_{j} V + V Q^{a T}_{j}) - i g^{\prime} B_{\mu}(Y V + V Y), 
\eeqa{cov_der}
where $W^a_j$ ($a=1\ldots3$) and $B$ are the $SU(2)$ and $U(1)$ gauge 
fields, respectively,
and $g_j$ and $g^\prime$ are the corresponding gauge couplings.  The 
generators are given by $Y =$ diag $(-1/2, -1/2, 0, 1/2, 1/2)$ and    
\beq
Q_L^a = \left( \begin{array}{ccc} \tau^a& & \\ & & \\ & & \end{array}
\right),~~~~Q_R^a = \left( \begin{array}{ccc} & & \\ & & \\ & &-\tau^{a*} 
\end{array} \right). 
\eeq{gaugedgens}    
We will assume that the left-handed fermions of the Standard Model transform
as doublets under $SU(2)_L$ and singlets under $SU(2)_R$. 
\subsection{Vector boson mass matrices}

From this starting point, it is not difficult to work out the masses and 
couplings of the vector bosons in this theory and compute their effect on 
the precision electroweak observables.  In the basis $(A_L^+, A_R^+)$, the 
mass matrix of charged vector bosons is
\beq
    m_+^2 =  {f^2\over 2} \pmatrix{ g_L^2     &    -  \half g_Lg_R (1+c) \cr
                                          -\half g_Lg_R (1+c) &    g_R^2\cr}.
\eeq{firstmW}
The mass of the heavy $W$ gauge bosons to leading order is again given by
Equation~(\ref{Wheavythree}).  Finding the masses of the usual $W$ and $Z$ to 
the precision that is necessary to compare with electroweak 
precision measurements requires a bit more work.
Let
\beq
      g^2 =  {g_L^2 g_R^2\over g_L^2 + g_R^2} \ .
\eeq{gdef}
We can define a mixing angle $\psi$ by 
\beq
        g_L =  {g\over c_\psi} \ , \qquad    g_R = {g\over s_\psi},
\eeq{gLgRdef}
where $s_\psi\equiv\sin\psi$, $c_\psi\equiv\cos\psi$. In the basis 
\beq
A_{(-)} = s_{\psi} A_{L} - c_{\psi} A_{R} 
\ , \qquad A_{(+)} = c_{\psi} A_{L} + s_{\psi} A_{R} 
\eeq{bigrot}
the matrix $m_+^2$ is approximately diagonal. A further rotation 
of order $v^{2}/f^{2}$ is necessary to complete the diagonalization. 
The mass eigenstates are given by
\beq W^{+} =  s_\beta A_{(-)}^{+} + c_\beta A_{(+)}^{+} \ , \qquad  
 W_H^{+} =  c_\beta A_{(-)}^{+} - s_\beta A_{(+)}^{+} \ ,
\eeq{betadef}
where 
\beq
s_\beta \approx {v^2\over 4f^2}\,c_\psi s_\psi
        (c_\psi^2 - s_\psi^2), \qquad  c_{\beta} \approx 1.
\eeq{betavalue}
Here and below, we neglect the terms of order $v^4/f^4$ and higher. 
The $W_H$ boson receives a mass of order $f\sim$ TeV, while the $W$ boson 
remains light. Its mass is given by
\beq
    m_W^2 =    {g^2 v^2 \over 4} \left[ 1 -  {v^2\over f^2}
           \left( {1\over 12} + {1\over 8} (c_\psi^2 - s_\psi^2)^2\right)
                                                  +  \ldots \right].
\eeq{mWfinal}
The effective value of $G_F$, including the effect of the exchange of both 
vector bosons at $Q^2 = 0$, is
\beqa
    { G_F\over \sqrt{2}} &= &{1\over 8}  \pmatrix{g_L & 0 \cr}   m_+^{-2} 
                  \pmatrix{g_L \cr 0 \cr} \CR
    &=& {1\over 2v^2} \left[ 1 + {5\over 24} {v^2\over f^2}
                             \right] \ .
\eeqa{GFvalues}

Similarly, the mass matrix of neutral vector bosons in the basis $(A_L^3,
A_R^3,B)$ is given by
\beq
    m_0^2 =  {f^2\over 2} \pmatrix{ 
g_L^2  (1+\zeta)    & - g_Lg_R(\half (1+c)+\zeta) & -\half g_L g' (1-c)\cr
- g_Lg_R(\half (1+c)+\zeta) &    g_R^2(1+\zeta)   &   - \half g_R g' (1-c)\cr
- \half g_L g' (1-c) &    - \half g_R g'(1-c)     &   g^{\prime 2}(1-c) \cr} 
\ ,
\eeq{firstmZ}
where $\zeta = (1-c)^2/8$. Comparing to~\leqn{firstmW}, we see that the 
terms proportional to $\zeta$ in the  matrix elements violate custodial 
$SU(2)$; however, these terms only contribute to $\mw/\mz$ in order $v^6/f^6$. 
Let $\theta_u$ denote the ``underlying'' value of the weak mixing angle, 
defined by
\beq
       g =  {e\over s_u} \ , \qquad g' =  {e\over c_u} \ , \qquad 
\eeq{sudef}
where $s_u\equiv\sin\theta_u$, $c_u\equiv\cos\theta_u$, and 
$e=gg^\prime/\sqrt{g^2+g^{\prime 2}}$. We can now proceed to the new basis:
\beq
A_{(-)}^3, \qquad 
Z_{(0)} = c_{u} A^3_{(+)} - s_{u} B, \qquad 
A = s_u A^3_{(+)} + c_u B,
\eeq{newbasis}
where $A_{(+)}$ and $A_{(-)}$ are defined in Equation~\leqn{bigrot}. The state
$A$ is an exact eigenvector of $m_0^2$ with a vanishing eigenvalue; we 
identify this state with the physical photon. The other two states in 
Equation~\leqn{newbasis} are not exact eigenvectors. As in the charged sector,
a further rotation of order $v^{2}/f^{2}$ is needed to complete the 
diagonalization:  
\beq 
Z =  s_{Z} A^3_{(-)} + c_{Z} Z_{(0)} \ , \qquad
   Z_H =  c_{Z} A^3_{(-)} - s_{Z} Z_{(0)} \ ,
\eeq{Zeigens}
where 
\beq
   s_{Z} \approx {v^2\over 4f^2}{c_\psi s_\psi\over c_u}
        (c_\psi^2 - s_\psi^2), \qquad  c_{Z} \approx 1.
\eeq{gammavalue}
The mass of the light $Z$ boson is given by
\beq
    m_Z^2 =    {g^2 v^2 \over 4c_u^2} \left[ 1 -  {v^2\over f^2}
           \left( {1\over 12} + {1\over 8} (c_\psi^2 - s_\psi^2)^2\right)
+  \ldots \right],
\eeq{mZfinal}
while the $Z_H$ state obtains a mass of order $f\sim$ TeV. 

\subsection{Precision electroweak observables}

From these formulae, we can work out the predictions for corrections to 
precision electroweak observables due to heavy gauge bosons.
The reference value $\theta_0$ of the weak mixing angle is 
given by
\beq
     \sin^2 2\theta_0 = { 4\pi \alpha \over \sqrt{2} G_F m_Z^2}.
\eeq{refvalofsstw}
From the formulae in Equations~(\ref{GFvalues}), (\ref{mZfinal}) 
and (\ref{refvalofsstw}), we can compute the shift between the underlying
$s_{u}^{2}$ of Equation~(\ref{sudef}) and the reference value of 
$\sin^{2} \theta_{0}$, defined above:
\beq
     s_u^2 =  s_0^2 +\Delta s^2\ ,
\eeq{suval}
with
\beq
\Delta s^2 =  \half {v^2\over f^2} c^2_\psi s^2_\psi 
      {c_0^2s_0^2\over c_0^2-s_0^2}.   
\eeq{valueofsu}
Here, we have defined $s_0^2 \equiv \sin^2\theta_0, c_0^2=1-s_0^2$.   

Using this formula and the expression for $s_{\beta}$ in 
Equation~(\ref{betavalue}) to compute the coupling
of leptons to the $Z^0$, we can compute the shifts of precision electroweak 
observables from their Standard Model tree-level 
values. For the three
best-measured observables---$\mw$, the on-shell mass of the $W$, $s_*^2$, 
the effective value of the 
weak mixing angle in $Z^0$ decay asymmetries, and $\Gamma_\ell$, the leptonic
width of the $Z^0$, we find
\beqa
  \Delta m_W & \equiv & \mw - \mz c_0 =  
                 - \half{\mw \over c_0^2} \Delta s^2, \CR
 \Delta   s_*^2 & \equiv &  s_*^2 - s_0^2  =   \Delta s^2
 - {1\over 4}{v^2\over f^2}s_0^2 s_\psi^2(c_\psi^2-s_\psi^2),   \CR
 \Delta \Gamma_\ell & \equiv & \Gamma_\ell - \Gamma_{\ell 0} = 
 -  \Gamma_{\ell 0} \left[ \half {v^2\over f^2} s^4_\psi +  {4 (1-4s_0^2)\over 
         1 - 4s_0^2 + 8 s_0^4} \Delta s^2 \right],
\eeqa{DeltasofEW}
where 
\beq
    \Gamma_{\ell 0} = {4\pi e^2\mz\over 6  s_0^2 c_0^2}
               \left( (\half - s_0^2)^2 + s_0^4 \right) 
\eeq{Gammaellzero} 
is the Standard Model tree-level value of the leptonic width of the $Z$. 

We can interpret these shifts as a contribution to the $S$ and $T$ 
parameters~\cite{PT}.
Formally, effects on the precision electroweak parameters due to extra $Z$
and $W$ bosons  are not oblique and cannot be completely absorbed into 
$S$ and $T$.  However, it was observed in \cite{PandW} that a fit to the 
electroweak data with the shifts from a $Z'$ and compensatory values of 
$S$ and $T$ was comparable in quality to a fit to the Standard Model; the 
opposite of the 
values of $S$ and $T$ needed to compensate the effect of the $Z'$ could then
be viewed as the $(S,T)$ excursion due to the $Z'$. 

\begin{figure}[th!]
\begin{center}
\epsfig{file=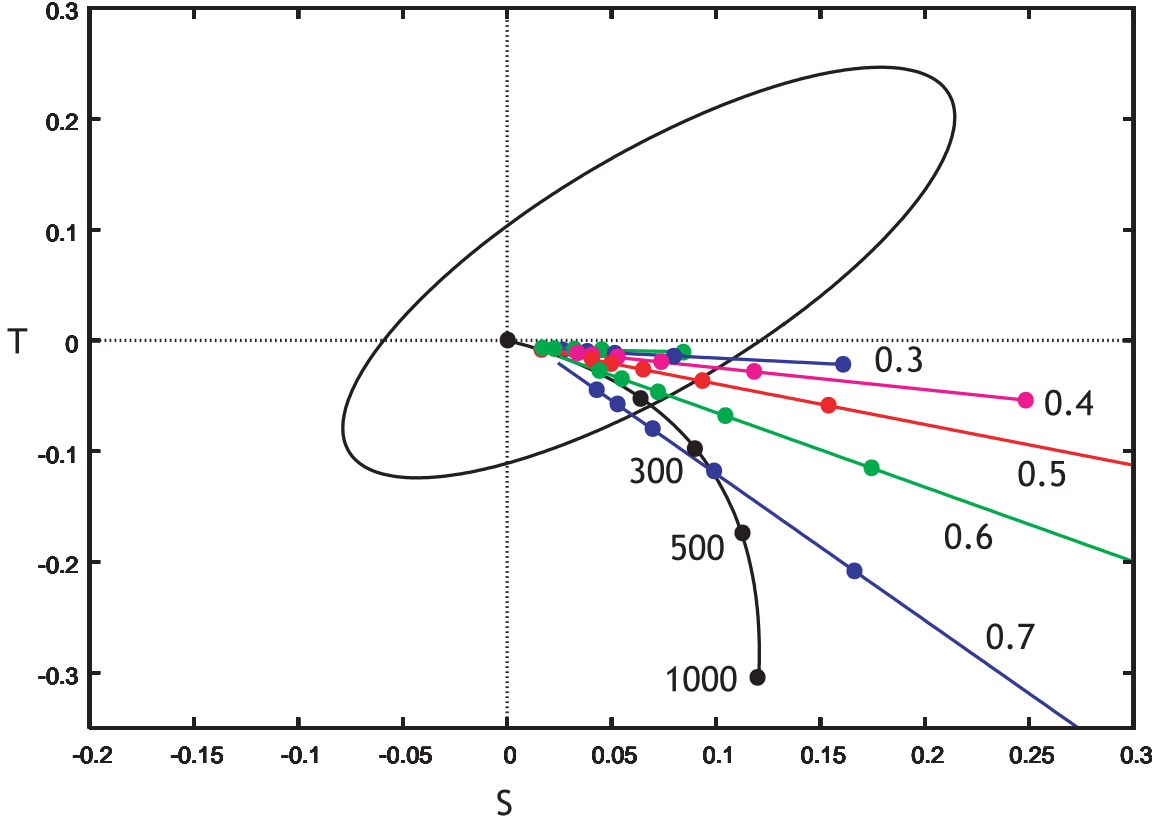,width=\columnwidth}
\caption{Excursions in the $S-T$ plane resulting from the $SU(5)/SO(5)$
Little Higgs model with a single gauged $U(1)$.  The different lines 
represent different 
values of $\sin \psi$=\{0.1, 0.2, 0.3, \ldots 0.7 \}, while the points
on the lines represent different values of $f$.   The rightmost point 
(not visible for $\sin \psi=\{0.5,0.6,0.7\}$) is for
 1 TeV, and additional points are separated by 500 GeV, increasing in $f$. 
The ellipse represents the 
experimentally allowed region at the  68\% confidence level for two degrees 
of freedom.  Also shown is the dark black 
curve showing the $S$ and $T$ contributions of a Standard Model Higgs boson 
for various masses.  We provide an enlargement of this figure 
in Fig.~\ref{fig:STbig}.  Note that the lines for 
$\sin \psi$=\{0.1, 0.2\} are somewhat obscured here, but can be seen clearly
in Fig.~\ref{fig:STbig}.}
\label{fig:ST}
\end{center}
\end{figure}

\begin{figure}[th!]
\begin{center}
\epsfig{file=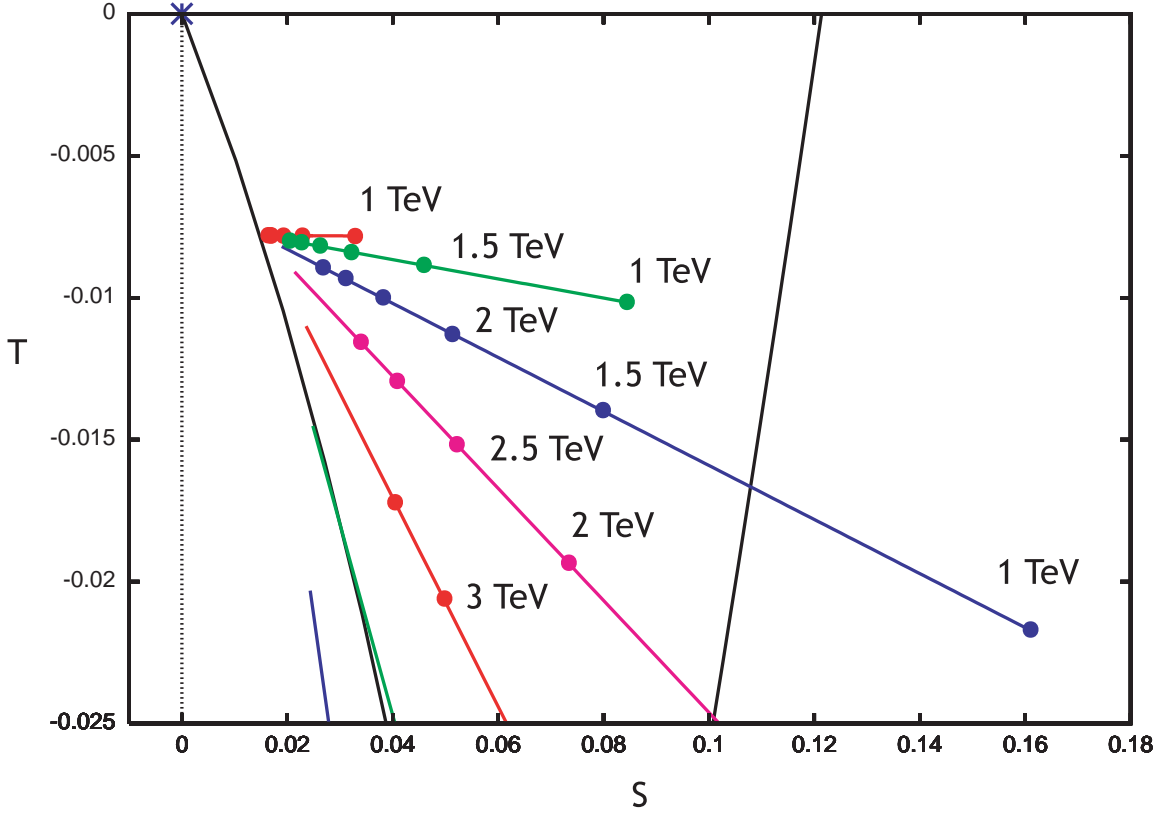, width=\columnwidth}
\caption{This figure is an
enlargement of the central portion of Fig.~\ref{fig:ST}, focusing on 
points compatible 
with the electroweak precision measurements.  The uppermost line is for 
$\sin \psi$=0.1.  Each line represents an increment of 0.1 in $\sin \psi$.
The Higgs mass is set to $115$ GeV in the Little Higgs model, 
while $S=T=0$ is defined for $m_{h}$=100 GeV.}
\label{fig:STbig}
\end{center}
\end{figure}

Applying this method to the $SU(5)/SO(5)$  Little Higgs model with 
$SU(2)\times SU(2) \times U(1)$ gauged, using the observables 
in~\leqn{DeltasofEW}, we find that the effect of the model on the precision 
electroweak data is represented by the $(S,T)$ excursions shown in 
Fig.~\ref{fig:ST} and Fig.~\ref{fig:STbig}.  In producing the fit, we 
use \cite{EWPData}
\beqa
m_{W}&=& 80.425 \pm 0.034 \mbox{ GeV}, \\
s_{\ast}^{2}&=& 0.23150 \pm .00016,\\
\Gamma_{\ell}&=& 83.984 \pm 0.086 \mbox{ MeV}.
\eeqa{kgbdk}
We allow the values of the top quark mass and the electromagnetic coupling
to vary within their current errors: $\mt = 174.3 \pm 5.1$ GeV, 
$\alpha^{-1} (\mz) = 128.936 \pm 0.021$~\cite{EWPData}.  
  All effects shown 
in~\leqn{DeltasofEW} become very small as $s_\psi\to 0$, allowing lower 
values of $f$ to be consistent with the electroweak constraints. The 
constraint cannot be eliminated completely, since according to
Equation~\leqn{gLgRdef} the gauge coupling $g_R$ becomes strong 
as $s_\psi \to 0$
and the perturbative analysis performed here is no longer applicable.
Still, the bounds on $f$ are not very strong: for example, for $s_\psi = 0.2$ 
(corresponding to $g_R^2/4\pi \approx 0.4$, which is probably not yet strong 
coupling), we find that $f$ can be lower than 1 TeV within the 68\% 
confidence region of the electroweak fit. 

Our analysis includes the shifts of the electroweak precision observable 
due to heavy 
gauge bosons, but does not include possible corrections from a vacuum 
expectation value (vev) for the $SU(2)$ triplet pseudo-Goldstone boson present 
in this model.  These corrections can play 
an important role in constraining the model for small values of 
$s_\psi$~\cite{Csaki2}. The value of the triplet vev is not calculable from
the low-energy effective Little Higgs theory, and including it in the
analysis corresponds to adding an extra free parameter to the model. 
The bounds in Fig.~\ref{fig:ST} and Fig.~\ref{fig:STbig} are valid in the 
regions of the parameter space where the triplet vev is small, but may
underestimate the constraints in other regions. A more detailed analysis
that includes the contribution of the triplet vev can be found in 
Ref.~\cite{Csaki2}.
  
Since all three of the measurements in \leqn{DeltasofEW} are made at the
$Z$ and $W$ poles, one should ask whether low-$Q^2$ observables can put
further constraints on the parameters of the Little Higgs theory.  To analyze 
this, we have computed the effective low-energy neutral current Lagrangian, 
in the form
\beq
  \mathcal{L}_{\rm\tt NC} = {G_F\over\sqrt{2}} 
\rho\left[(J^3_\mu - s_\nu^2 J^Q_\mu)^2 
                                 + \eta  J^Q_\mu J^{Q\mu} \right] \ .
\eeq{effNC}
In the Standard Model at tree level, $\rho = 1$, $s_\nu^2 = s_0^2$, $\eta = 0$.
In the $SU(5)/SO(5)$ model, we find $\rho = 1$, up to corrections of order
$v^6/f^6$, and 
\beqa
     s_\nu^2 -s_0^2 &=&   s_0^2  \cdot {1\over 4}{v^2\over f^2} s_\psi^2
             \left[ {2c_0^2 c_\psi^2\over c_0^2 -s_0^2} - 1\right],     \CR
      \eta &=&   s_0^4\cdot  \half {v^2\over f^2} s^4_\psi \ .
\eeqa{snuandeta}
For points in the region allowed by the $(S,T)$ analysis, the shifts in 
$s_\nu^2$ are very small.  For example, for $s_\psi^2 = 0.2$ and $f = 1$ 
TeV, $s_\nu^2 - s_0^2 \approx 10^{-4}$.  The parameter $Q_W$ of atomic parity 
violation~\cite{APV} and the observables $R^\nu$, $R^{\bar \nu}$ measured by 
the NuTeV experiment~\cite{Nutev} depend on $s_\nu^2$ but do not 
involve $\eta$.  In all
cases, the effects on these parameters are corrections of relative size less 
than $10^{-3}$, well within the current experimental errors.

Through \leqn{mTlower}, the lower bound on 
$f$ from the precision electroweak observables
places a strong lower bound of about 2 TeV on the mass of the $T$.    
However, this still leaves a range in which the $T$ can be discovered at the 
LHC.  It is worth emphasizing that a $T$ mass much higher than 2 TeV would
imply a large amount of fine tuning in the Higgs potential. Therefore, 
naturalness considerations together with precision electroweak constraints 
indicate that if the Little Higgs model is correct, the heavy top should be in 
the 2 TeV range. In this case, it is possible that the physics of electroweak 
symmetry breaking in the Little Higgs model can be tested at LHC.  We
now turn to the analysis of those experimental tests.  For our analyses in 
the next section, we assume a heavy top mass of 2.5 TeV and $f$=1.2 TeV, 
which is clearly allowed by the precision electroweak observables.

\section{Testing the Model at the LHC}

To test the relation \leqn{Tmasstwo}, it is necessary to measure three 
quantities, the parameter $f$, the mass $m_T$, and the coupling constant 
$\lambda_T$. The measurement of the mass and production cross section of 
the heavy $SU(2)$ gauge bosons $W_H, Z_H$ at the LHC can be used to 
determine $f$~\cite{AaronMaxim}. We will review the strategy for this 
measurement below, concentrating on the low values of the mixing 
angle, $\psi$, preferred by precision electroweak constraints. The 
measurement of the heavy top mass $m_T$ is rather straightforward; on 
the other hand, it is much less clear how $\lambda_T$ can be 
determined.  In this section, 
we will discuss two methods for measuring $\lambda_T$.  These involve 
the decay width and the production cross section for $T$ quarks 
at the LHC.

\subsection{Measuring the parameter $f$} \label{sec:getf}

In the $SU(5)/SO(5)$ Little Higgs model described in 
section~\ref{PEW}, all the couplings involving the heavy gauge 
bosons $W_H^\pm$ and $Z_H$ depend on just two unknown parameters, the 
scale $f$ and the mixing angle $\psi$, defined in Equation~(\ref{gLgRdef}). 
Thus, a small number of measurements in this sector is sufficient to 
determine both parameters. Let us concentrate on the measurements 
involving the neutral gauge boson $Z_H$.  To leading order in $v/f$, the 
$Z_{H}$ mass is given by
\beq
M_{Z_{H}} = \sqrt{\frac{g_{L}^{2} 
        +g_{R}^{2}}{2}} f = \frac{\sqrt{2} g}{\sin 2\psi}\,f.
\eeq{massZh}
The production cross section and decay branching ratios of $Z_H$ 
bosons have been obtained\footnote{The conventions used in 
Ref.~\cite{AaronMaxim} are slightly different from the ones used in this 
paper; they are related by
$f_{\cite{AaronMaxim}}=\sqrt{2}f_{\rm \tt here}$, $\psi_{\cite{AaronMaxim}}=
\pi/2-\psi_{\rm \tt here}$.} in~\cite{AaronMaxim}. For fixed $M_{Z_{H}}$, the 
production cross section is proportional to $\tan^2\psi$. The decay rate is 
given by
\beq
\Gamma = \frac{g^2}{96\pi}\,(\cot^2 2\psi + 24 \tan^2 \psi)\,M_{Z_{H}}\,,
\eeq{gammaZh}
with the branching ratios\footnote{Here we correct a mistake 
in Ref.~\cite{AaronMaxim}, where the $W^+W^-$ decay mode was 
inadvertently omitted~\cite{Heather}.}
\beqa
\Br(\ell\bar{\ell}) &=& \frac{1}{3}\,\Br(q\bar{q}) = 
\frac{\tan^2\psi}{\cot^2 2\psi + 24 \tan^2 \psi}; \CR
\Br(W^+W^-) &=& \Br(Zh) = \frac{1}{2}\,\frac{\cot^2 2\psi}{\cot^2 2\psi + 24 
\tan^2 \psi}.
\eeqa{brZh} 
From these formulae, it is clear that combining, for example, the 
measurement of the $Z_H$ mass and the number of events in the 
$\ell^+ \ell^-$ ($\ell=e$ or $\mu$) channels is sufficient to 
determine both $f$ and $\psi$. 

In the parameter region preferred by electroweak precision constraints, 
the dominant decay modes are $Z_H\to W^+ W^-$ and $Z_H\to Zh$. For example,
for $s_\psi=0.2$, the combined branching ratio of these two modes is 
about 85\%, with the remaining decays to fermion pairs.  The branching 
ratio to leptons ($e$'s and $\mu$'s) is only
about 1\%.  Nevertheless, for an $f=1.2$ TeV the production cross section 
for the $Z_H$ is roughly 12 fb, corresponding to 3600 events in a 
300 fb$^{-1}$ data sample.  Therefore, in the lepton channels we still 
expect roughly 40 events, with virtually no background. Studying these
events should be sufficient to determine $f$ and $\psi$. Of course, the 
events in the other decay channels, along with the decays of $W_H^{\pm}$, 
will only help to improve the precision of the determination of $f$.

\subsection{Measuring $\lambda_{T}$}
\subsubsection{Decays of the $T$ quark}

Since $T$ has a vertex for $T\to t h$, as shown in Fig.~\ref{fig:zerorules}, 
the heavy $T$ quark will decay to $th$, and the corresponding decay width
is proportional to $\lambda_T^{2}$.  But $T$ also has other decay modes.  This
is made clear by looking at the `gaugeless limit'~\cite{bj}  $g\to 0$, in 
which the weak bosons become massless and the Goldstone bosons of $SU(2)
\times U(1)$ breaking become physical.  In this limit, the structure of 
\leqn{thedelta} ensures that
$T$ decays symmetrically to the four members of the Higgs $SU(2)$ doublet:
$\Gamma (T \to t h) = \Gamma(T\to t \pi^3) =  \half \Gamma(T\to b \pi^+)$.
In the real situation, $\pi^+$ and $\pi^3$ are replaced by the longitudinal 
polarization states of the $W^+$ and $Z^0$ vector bosons:
\beq
\Gamma (T \to t h) \approx \Gamma(T\to t Z^0) \approx  \half \Gamma(T\to 
b W^+)\ .
\eeq{GammaTrel}
All three decay modes provide characteristic signatures for the discovery of 
the $T$ at the LHC.

We will now obtain more exact relations for the decay branching ratios of the 
$T$ and, at the same time, see how the approximate equalities \leqn{GammaTrel}
work when the Standard Model gauge couplings are turned back on.
To do this, we must diagonalize the top quark mass
matrix more carefully, picking up terms that we dropped in the discussion
leading to \leqn{tcouplings}.  In principle, we should also 
modify \leqn{thedelta} to 
take into account the constraints from using an $SU(5)/SO(5)$ nonlinear
sigma model.  However, this model belongs to the general class of models for
which the formulae of Section 2 are precisely valid.  To see this explicitly,
the invariant Lagrangian can be written in terms of $SU(5)/SO(5)$ 
Goldstone bosons as~\cite{LH2}
\beq
  \mathcal{L} =   - {\lambda_1\over 2}f\, \bar u_R 
                      \epsilon_{ijk} \epsilon_{mn} V_{im} 
V_{jn} \chi_{L k} - \lambda_2 f\, \bar U_R U_L + {\rm h.c.}\  ,
\eeq{thedeltafive}
where $V_{im}$ denotes the $3\times 2$ upper right hand block of $V$ in 
\leqn{theVforfive}.  The relevant Feynman rules for the top quarks are
 just those shown in Fig.~1.  According to Equation~(\ref{firstmW}), $f$ is 
again given in terms of the heavy boson masses by 
Equation~(\ref{Wheavythree}), and is fixed to be roughly greater than 
1 TeV by the arguments of the previous section. 

Now let us consider the heavy quark mass diagonalization more carefully.
 In particular, if we include the $SU(2)\times
U(1)$-breaking vacuum expectation value $v$, the top quark mass matrix
becomes
\beq
     \pmatrix{ \bar u_R & \bar U_R\cr} m_U      \pmatrix{u_L \cr U_L \cr}, \  
\eeq{mTrealdef}
with
\beq
    m_U = f
\pmatrix{ \frac{\lambda_1 s}{\sqrt{2}} &  \frac{\lambda_1}{2}(1+c) \cr
       0      &  \lambda_2  \cr} \   .
\eeq{realUmassmatrix}
Here, we have again used $s \equiv \sin \frac{v}{f}$ and 
$c \equiv \cos \frac{v}{f}$.
Diagonalizing $m_U^{\dagger} m_U$, we find the mixing angle for the 
left-handed components of the top quarks,
\beq
      \theta_t = \frac{1}{2} \tan^{-1}  
       \frac{ 2 \sqrt{2} \lambda_1^{2} s (1+c)}{4 \lambda_{2}^{2} 
      + (1+c)^{2} \lambda_{1}^{2} 
          - 2 \lambda_{1}^{2} s^{2}} \approx  {1\over \sqrt{2}}
      {\lambda_1^2 \over \lambda_1^2 + \lambda_2^2} {v \over f}\ 
               \approx {\lambda _{T} v \over \sqrt{2} m_{T}} 
\eeq{sinthetaval}
Let $\cos\theta_t = c_t$, $\sin\theta_t = s_t$; then the mass eigenstates are 
given by
\beqa
            T_L &=&    c_t  U_L + s_t u_L, \CR
          t_L &=&   - s_t  U_L + c_t u_L    \ .
\eeqa{thetatdef}
There is also a mixing angle for the right-handed quarks, $\theta_{tr}$, given
by
\beqa
            T_R &=&    c_{tr}  U_R + s_{tr} u_R, \CR
          t_R &=&   - s_{tr}  U_R + c_{tr} u_R    \ .
\eeqa{thetatRdef}
Note that this angle is non-zero even in the absence of electroweak symmetry 
breaking, see Equation~(\ref{teigenstates}).

From the mixing, the top quark mass receives a small correction,
\beq
     m_t =   {\lambda_t v\over \sqrt{2}}  \left( 1 - 
           \left(\frac{1}{6} - \frac{\lambda_1^2 \lambda_2^{2}}{4 
 (\lambda_{1}^{2} +\lambda_{2}^{2})^{2}}\right) {v^2\over f^2} 
               + \cdots \right) \ .
\eeq{mtcorrect}
For $f = 1.2$ TeV, this is a  0.3\% correction to the Standard Model
tree level relation.   In the following, we will quote `exact' tree-level 
relations 
in terms of $m_t$, $m_T$, $\lambda_T$, and $\theta_t$ and their leading-order
terms in an expansion in $v/f$.  Typically, these expressions will agree to 
within a 
few percent. 

The admixture of $u_L$ in the $T$ allows this quark to decay by the Standard
Model weak currents.  The amplitudes for the decay modes to $W^+$ and $Z^0$
are then proportional to $\sin\theta_t$.  However, the contraction of the 
longitudinal polarization vector of a massive vector boson with the 
spontaneously broken weak current gives an enhancement by a factor 
$m_T/\mw$, so that the full coupling is of the order of
\beq
          {g  \over \sqrt{2}} \left({m_T\over m_W} \right) \theta_t  =  
                \sqrt{2} {m_T\over v} \theta_{t} 
                         =   \lambda_T    \ .
\eeq{recoverlambdaT}
This allows the three branching fractions of the $T$ to be of the same order
of magnitude.  A similar effect is well known in the decays of the singlet
$D$ quark in $E_6$ models~\cite{Esix}.  

Working more explicitly, we find for the three dominant partial widths 
of the $T$ quark: 
\beqa
\Gamma(T\to t h) &=&   {\lambda_{1}^{2}  \, m_{T}  \over 64 \pi} 
f(x_{t}, x_{h})[(1 + x_{t}^{2} - x_{h}^{2})(C_{L}^{2} + C_{R}^{2})
+ 4 C_{L} C_{R} x_{t}] \CR &\approx& {m_{T} \lambda_{T}^{2} \over 
64 \pi}, \CR
\Gamma(T\to t Z^0) &=&  {e^{2} \sin^{2} 2 \theta_{t} m_{T}^{3} c_{Z}^{2} 
\over 512 \pi M_{Z}^{2} s_{u}^{2} c_{u}^{2}}  (1 + \tan_{\psi} c_{u} 
\tan_Z)^{2} \, f(x_{Z},x_{b}) \, g(x_{b},x_{Z})  \CR
&\approx&  {m_{T} \lambda_{T}^{2} \over 64 \pi} ,\CR
\Gamma (T\to b W^+)&=& {g^{2} \sin^{2} \theta_{t} m_{T}^{3} c_{\beta}^{2}
\over 64 \pi M_{W}^{2}} (1 + \tan_{\psi} \tan_{\beta})^{2} \, f(x_{W},x_{b}) 
\, g(x_{b},x_{W}) \CR 
&\approx& {m_{T} \lambda_{T}^{2} \over 32 \pi}.
\eeqa{threeGammas}
Note that $s_{Z}$, defined in Equation~(\ref{Zeigens}), and $s_{\beta}$,
defined in Equation~(\ref{betadef}), are both of order $v^{2}/f^{2}$. We have 
defined $x_{i} \equiv m_{i}/m_{T}$, the kinematic functions
\beqa
f(x_{i},x_{j})&=&\sqrt{(1-(x_{i}+x_{j})^{2})(1-(x_{i}-x_{j})^{2})}, \CR
g(x_{i},x_{j})&=&(1-x_{i}^{2})+x_{j}^{2} (1+x_{i}^{2}) -2 x_{j}^{4},
\eeqa{kinematics}
and introduced the couplings
\beqa
C_{L}=c_{tr} ( s_{t} \cos{\frac{v}{f}}  - {c_{t} \over \sqrt{2}} 
        \sin{\frac{v}{f}}) 
        \sim \orderof{{v \over f}}, \CR
C_{R}=s_{tr} (c_{t} \cos{\frac{v}{f}} + {s_{t} \over \sqrt{2}} 
        \sin{\frac{v}{f}}) \sim \orderof{1}. 
\eeqa{Tcouplings}

To leading order in $v^{2}/f^{2}$, the total width of the $T$ is then
\beq
     \Gamma_T = {m_{T} \lambda_{T}^{2} \over 16 \pi}.
\eeq{totalGT}
The measurement of this total width is the first possible method for 
measuring $\lambda_T$.  However, it is not so easy to measure the width of a 
strongly interacting particle produced at a hadron collider, because the 
fluctuations of QCD jets lead to an intrinsic smearing of the mass peak.
For the ATLAS detector, the fractional uncertainty in the two-jet 
invariant mass at 2.5 TeV is expected to be about $\pm 5$\%~\cite{askIan}, 
which for a heavy top of $2.5$ TeV, corresponds to a minimal error of 
$\pm 125$ GeV in the width.   
On the other hand, for $m_T = 2.5$ TeV and $\lambda_T 
\approx \lambda_t$, the formula~\leqn{totalGT} evaluates to only 50 GeV. 
With these estimates, $\Gamma_{T}$ will be only marginally visible, and then 
only if the jet mass resolution is very well understood theoretically.  
Therefore,  it appears challenging to use this strategy to make the test of 
the Little Higgs model described here. 

\subsubsection{Production of the $T$ quark}

\begin{figure}[th!]
\begin{center}
\epsfig{file=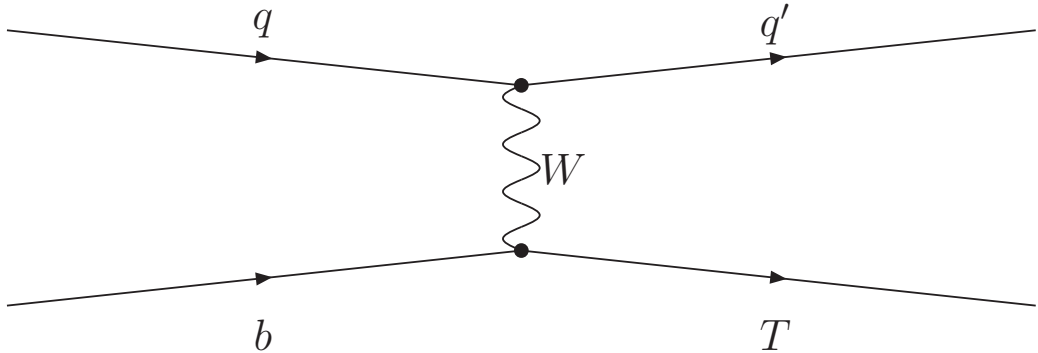,width=\columnwidth}
\caption{The dominant process for $T$ production at the LHC.}
\label{fig:makeaT}
\end{center}
\end{figure}

Another possible strategy is to extract $\lambda_T$ from a measurement of 
the $T$ production cross section at the LHC. For $m_T$ above 2 TeV, energy is 
at a premium and the single-$T$ production reaction $pp\to T+X$ dominates 
over the pair-production of $T$'s via strong interactions, $pp\to T+\bar{T}$. 
At the parton level, the dominant mechanism of single-$T$ production is 
through the ``$W-b$ fusion'' reaction~\cite{Han}
\beq
                q b \to   q^{\prime} T,
\eeq{makeaT}
shown in Fig.~\ref{fig:makeaT}. 
The cross section for this reaction is dominated by the exchange of a 
longitudinal $W$ boson, whose coupling to $T$ is proportional to $\lambda_T$ 
(see Equation~\leqn{recoverlambdaT}). Thus, a measurement of this cross section
would determine the value of $\lambda_T$, providing a test of the crucial
relation~\leqn{Tmasstwo}. 

\begin{figure}[th!]
\begin{center}
\epsfig{file=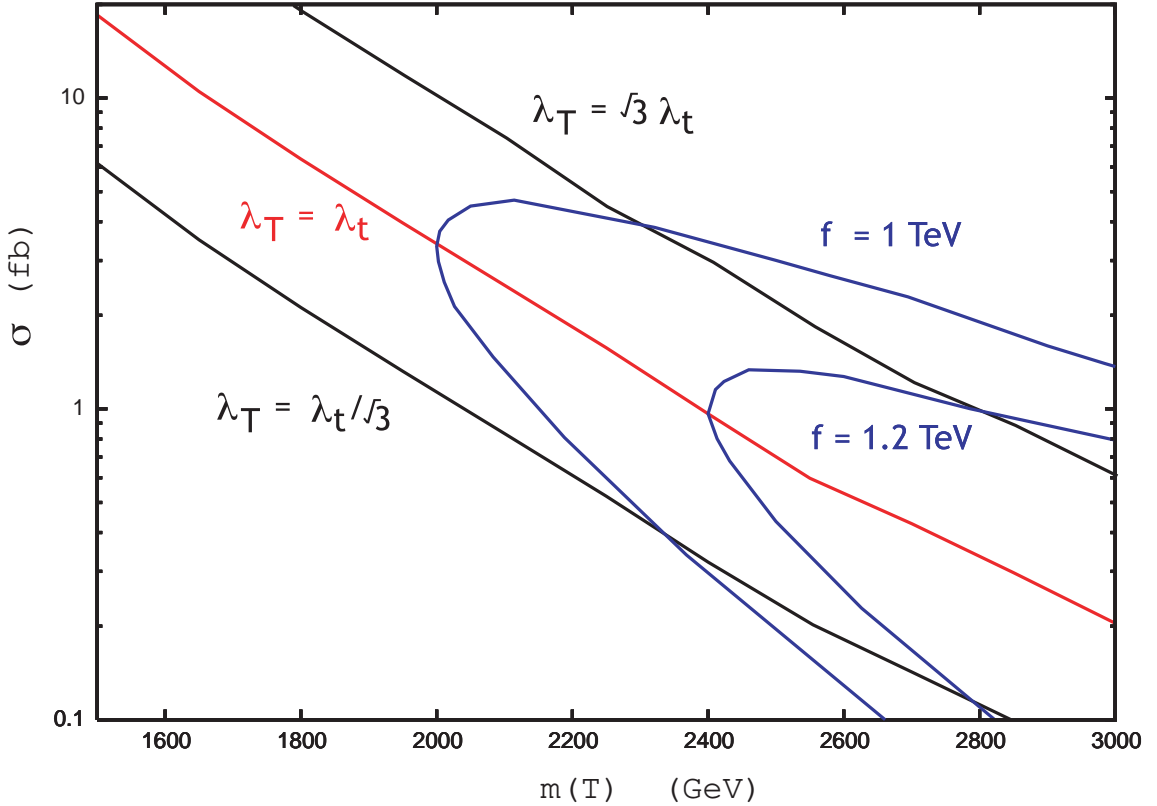,width=\columnwidth}
\caption{Parton level production cross section for the heavy top in the 
channel $b q \rightarrow T q\prime$ at the 14 TeV Large Hadron Collider.  The 
figure is made with the CTEQ4l parton distribution function.  The different 
lines show the difference in the production cross section for various 
values of $\lambda_{T}$.  The parabolas represent the predictions of the 
Little Higgs model for a constant $f$ as $\lambda_{T}$ is varied.}
\label{fig:TProd}
\end{center}
\end{figure}

How well can this cross section be measured at the LHC? The answer 
obviously depends on the mass of the $T$ quark, as well as on the size of
the coupling $\lambda_T$. In Figure~\ref{fig:TProd}, we plot the expected 
cross section as a function of $m_T$, for $\lambda_T=1/\sqrt{3}, 1$, and 
$\sqrt{3}$ using the CTEQ4l parton distribution functions. For  
$\lambda_{T}$ not too small, the number of events is large enough to
keep the statistical uncertainty under control: for example, for 
$m_T = 2.5$ TeV and $\lambda_{T}=1$, the evaluated production cross 
section corresponds to roughly 180 events for a 300 fb$^{-1}$ data sample. 
The reaction is 
characterized by a $T$ decay at low transverse momentum and in the central 
region, and all other jet activity very forward.  All three of the 
decay modes discussed above should be identifiable.   The final states
$T \rightarrow t h^{0} \rightarrow t b \bar{b}$ and 
$T \rightarrow b W^{+} \rightarrow b l^{+} \nu$ can be required with high 
efficiency and used to find a $T$ mass peak.  In the latter case, one 
should replace the observed $l^{+}$ with a $W^{+}$ in the $l^{+}$ direction.

Also shown in the figure are two parabolas, which represent the 
predictions of the model for two representative values of $f$.  Once 
the $f$ value is determined as described in Section~\ref{sec:getf}, the 
electroweak symmetry 
breaking mechanism described here predicts that the values of $m_T$ and
the production cross section lie on the corresponding parabola.

Converting a cross section measurement into a measurement of the coupling
$\lambda_T$ requires knowledge of the parton distribution functions (pdfs)
of the initial state particles. Since the typical energies involved are much  
larger than the $W$ boson mass, it is reasonable to use the effective-$W$
approximation, which treats the $W$ as a parton within the proton. In this
approximation, the single-$T$ production is described as a $2\to 1$ process,
$bW \to T$. The cross section is given by   
\beq
\sigma(pp\to T+X)\,=\,\int_0^1\,dx_b\,f_b(x_b,Q^{2})\,
f_w(\frac{m_T^2}{x_bS},Q^{2})\, \hat{\sigma}(Wb\to T),
\eeq{csT}
where $f_{b,w}$ are the pdfs of the $b$ quark and the $W$ boson, 
$\hat{\sigma}$ is the parton-level cross section, $S$ is the usual 
Mandelstam variable, and $Q$ is the renormalization scale, $Q^2\sim m_T^2$.   
The $b$ quark pdf is derived perturbatively from the gluon 
pdf~\cite{bpdf1,bpdf2,bpdf3}. The integral in~\leqn{csT} receives 
significant contributions from the region where 
$x\approx m_T/\sqrt{S}$. At the LHC, $\sqrt{S}=14$ TeV, and 
this region can extend to $x$ as high as $\sim 0.2$ for the values of 
the $T$ mass considered here. Currently, our knowledge of the $b$ pdf in the
large-$x$ region is rather poor: the uncertainty on $f(x_b)$ is 
about 20\% for $x_b = 0.1$ and even higher for higher $x_b$~\cite{Sullivan}. 
Without reducing this uncertainty, even a very accurate measurement of the 
single-$T$ production cross section would not provide a precision test of the 
relation~\leqn{Tmasstwo}. 

One possible way to reduce the uncertainty is to 
obtain an accurate measurement of the cross section of the Standard Model 
single top production at the Tevatron.  While there are several contributions 
to this process, the cross section is dominated by the 
$Wb$ fusion process, $Wb \to t$, and it has been shown~\cite{TimSep} that 
this contribution to the cross section can be isolated using kinematic cuts.  
A significant fraction of the events in this channel are initiated by $b$
quarks with $0.1 \lapproxeq x_b \lapproxeq 0.2$; the remaining events almost
exclusively come from the region $x_b < 0.1$, where the $b$ pdf is known much
more accurately. Thus, assuming that the value of $V_{tb}$ and the $W$ boson 
pdf are known, a measurement of $\sigma(p\bar{p}\to t+X)$ with the relevant 
cuts can be interpreted as a measurement of $f_b(x_b, Q^2)$ at 
$x_b\sim0.1-0.2$ and $Q^{2} \sim m_t^{2}$. This knowledge can then be used to 
reduce the uncertainty of the theoretical prediction for the $T$ quark
production at the LHC. Since 
$b$ is a sea quark, the $b$ pdfs in the proton and anti-proton are virtually 
identical, and the fact that the Tevatron is a $p\bar{p}$ collider introduces
no additional complications. Evolution of the $b$ pdf 
from $Q^{2} \sim m_t^{2}$ to $Q^{2} \sim m_T^{2}$ can be performed 
perturbatively.  It is well known that 
$f_{b}(x,Q^{2})$ decreases with increasing $Q^2$ at large $x$ and increases 
with $Q^{2}$ at small $x$. Interestingly, the cross over point for $Q^{2} 
\sim$ (1 TeV)$^{2}$ falls in the range of $x$ most relevant for the present 
discussion, $x \approx 0.18$.  For $0.14 <x < 0.2$, $f_{b}(x)$ varies by 
only a few percent going from $Q^{2}$=(175 GeV)$^{2}$ 
to $Q^{2}=$ (2 TeV)$^{2}$.  Therefore, 
measurements of $f_{b}(x,Q^{2})$ at the Tevatron can be extrapolated to the 
LHC with controllable uncertainty. The statistical uncertainty in the 
measurement
of $\sigma(p\bar{p}\to t+X)$ at the Tevatron is expected to be about 5\%
for 2 fb$^{-1}$ integrated luminosity~\cite{TY}. While a more detailed 
investigation is in order, it seems plausible that this method could result 
in a measurement of $\lambda_T^2$ at the level of 10\% or better.    

\section{Conclusions}
The Little Higgs model provides a simple mechanism for electroweak symmetry
breaking, relying on the top sector to trigger a negative (mass)$^{2}$ for the 
Higgs boson.  This mechanism depends on a simple relation between the 
parameters of the model, summarized in Equation~\leqn{Tmasstwo}.  In principle,
all parameters in this equation are measurable, providing a strong test of the
mechanism.  The testability of this relation at the LHC depends on the 
heavy top quark partner, $T$, being sufficiently light (roughly 2 TeV), as 
is favored by naturalness arguments.  We have shown that such a 
light $T$ can be consistent with current electroweak precision measurements.

The most challenging of the required measurements is the determination 
of  $\lambda_{T}$, the coupling of the heavy top to the Higgs boson.  We 
have outlined two strategies: measuring the width of the $T$ and 
measuring its production cross-section.  The first strategy, limited by 
calorimeter resolution, is not very promising.  The second strategy can be 
more successful if the pdfs of the $b$ quark can be determined more accurately 
at high $x$.  This may indeed be possible using the measurement of single 
top production from the current run at the Tevatron.

More detailed Monte Carlo studies to determine the feasibility of the 
measurements outlined here would be worthwhile. 

\Acknowledgements
We are grateful to Nima Arkani-Hamed, Gustavo Burdman, 
Csaba Csaki, Patrick Meade, Frank Petriello, Tim Tait, and 
Jay Wacker for valuable advice.
 MEP thanks Kiwoon Choi and his students for valuable 
discussions that ignited his interest in the topics considered here.
We thank Hitoshi Murayama for suggesting an improvement of our 
S,T fitting procedure.
M.~P. completed a part of this work at Lawrence Berkeley National 
Laboratory, where he was supported by the Director, Office of Science, 
Office of High Energy and Nuclear Physics, of the U.~S. Department of 
Energy under Contract DE-AC03-76SF00098.


\begin{thebibliography}{99}

\bibitem{LHorig}
N.~Arkani-Hamed, A.~G.~Cohen and H.~Georgi,
Phys.\ Lett.\ B {\bf 513}, 232 (2001)
[arXiv:hep-ph/0105239].

\bibitem{LHorig2}
N.~Arkani-Hamed, A.~G.~Cohen, T.~Gregoire and J.~G.~Wacker,
JHEP {\bf 0208}, 020 (2002)
[arXiv:hep-ph/0202089].

\bibitem{LH1}
N.~Arkani-Hamed, A.~G.~Cohen, E.~Katz, A.~E.~Nelson, 
T.~Gregoire and J.~G.~Wacker,
JHEP {\bf 0208}, 021 (2002)
[arXiv:hep-ph/0206020].

\bibitem{LH2}
N.~Arkani-Hamed, A.~G.~Cohen, E.~Katz and A.~E.~Nelson,
JHEP {\bf 0207}, 034 (2002)
[arXiv:hep-ph/0206021].

\bibitem{Schmaltz}
M.~Schmaltz,
Nucl.\ Phys.\ Proc.\ Suppl.\  {\bf 117}, 40 (2003)
[arXiv:hep-ph/0210415].

\bibitem{Wacker}
J.~G.~Wacker,
arXiv:hep-ph/0208235.

\bibitem{Georgi}
H.~Georgi and A.~Pais,
Phys.\ Rev.\ D {\bf 12}, 508 (1975).

\bibitem{GeorgiKaplan}
D.~B.~Kaplan and H.~Georgi,
Phys.\ Lett.\ B {\bf 136}, 183 (1984).

\bibitem{KGS}
D.~B.~Kaplan, H.~Georgi and S.~Dimopoulos,
Phys.\ Lett.\ B {\bf 136}, 187 (1984).

\bibitem{SUSYt1}
K.~Inoue, A.~Kakuto, H.~Komatsu and S.~Takeshita,
Prog.\ Theor.\ Phys.\  {\bf 68}, 927 (1982)
[Erratum-ibid.\  {\bf 70}, 330 (1983)].

\bibitem{SUSYt2}
L.~E.~Ibanez,
Nucl.\ Phys.\ B {\bf 218}, 514 (1983);
L.~E.~Ibanez and C.~Lopez,
Phys.\ Lett.\ B {\bf 126}, 54 (1983).

\bibitem{SUSYt3}
J.~R.~Ellis, J.~S.~Hagelin, D.~V.~Nanopoulos and K.~Tamvakis,
Phys.\ Lett.\ B {\bf 125}, 275 (1983).

\bibitem{SUSYt4}
L.~Alvarez-Gaume, J.~Polchinski and M.~B.~Wise,
Nucl.\ Phys.\ B {\bf 221}, 495 (1983).



\bibitem{topcolor}
C.~T.~Hill,
Phys.\ Lett.\ B {\bf 266}, 419 (1991),
Phys.\ Lett.\ B {\bf 345}, 483 (1995)
[arXiv:hep-ph/9411426].

\bibitem{Nelson}
A.~E.~Nelson,
arXiv:hep-ph/0304036.

\bibitem{Sp6}
I.~Low, W.~Skiba and D.~Smith,
Phys.\ Rev.\ D {\bf 66}, 072001 (2002)
[arXiv:hep-ph/0207243].

\bibitem{SK}
D.~E.~Kaplan and M.~Schmaltz,
arXiv:hep-ph/0302049.

\bibitem{Jay}
S.~Chang and J.~G.~Wacker,
arXiv:hep-ph/0303001.

\bibitem{Sp6Precision}
T.~Gregoire, D.~R.~Smith and J.~G.~Wacker,
arXiv:hep-ph/0305275.

\bibitem{Skiba:2003yf}
W.~Skiba and J.~Terning,
arXiv:hep-ph/0305302.

\bibitem{Spencer}
S.~Chang,
arXiv:hep-ph/0306034.


\bibitem{Csaki}
C.~Csaki, J.~Hubisz, G.~D.~Kribs, P.~Meade and J.~Terning,
Phys.\ Rev.\ D {\bf 67}, 115002 (2003)
[arXiv:hep-ph/0211124].

\bibitem{HPR}
J.~L.~Hewett, F.~J.~Petriello and T.~G.~Rizzo,
arXiv:hep-ph/0211218.


\bibitem{Csaki2}
C.~Csaki, J.~Hubisz, G.~D.~Kribs, P.~Meade and J.~Terning,
Phys.\ Rev.\ D {\bf 68}, 035009 (2003)
[arXiv:hep-ph/0303236].


\bibitem{AaronMaxim}
G.~Burdman, M.~Perelstein and A.~Pierce,
Phys.\ Rev.\ Lett.\  {\bf 90}, 241802 (2003)
[arXiv:hep-ph/0212228].

\bibitem{Han}
T.~Han, H.~E.~Logan, B.~McElrath and L.~T.~Wang,
Phys.\ Rev.\ D {\bf 67}, 095004 (2003)
[arXiv:hep-ph/0301040].

\bibitem{Han2}
T.~Han, H.~E.~Logan, B.~McElrath and L.~T.~Wang,
Phys.\ Lett.\ B {\bf 563}, 191 (2003)
[arXiv:hep-ph/0302188].

\bibitem{DHill}
B.~A.~Dobrescu and C.~T.~Hill,
Phys.\ Rev.\ Lett.\  {\bf 81}, 2634 (1998)
[arXiv:hep-ph/9712319];
R.~S.~Chivukula, B.~A.~Dobrescu, H.~Georgi and C.~T.~Hill,
Phys.\ Rev.\ D {\bf 59}, 075003 (1999)
[arXiv:hep-ph/9809470].




\bibitem{Jarry}
A. Jarry, {\it Ubu Roi}, Paris: Mercure de France, 1896.

\bibitem{EWWG}
\url{http://lepewwg.web.cern.ch/LEPEWWG/}

\bibitem{AHWsugg}
N. Arkani-Hamed and J. Wacker, personal communication.

\bibitem{GKSUtwo}
H.~Georgi and D.~B.~Kaplan,
Phys.\ Lett.\ B {\bf 145}, 216 (1984).

\bibitem{PT}
M.~E.~Peskin and T.~Takeuchi,
Phys.\ Rev.\ Lett.\  {\bf 65}, 964 (1990),
Phys.\ Rev.\ D {\bf 46}, 381 (1992).


\bibitem{PandW}
M.~E.~Peskin and J.~D.~Wells,
Phys.\ Rev.\ D {\bf 64}, 093003 (2001)
[arXiv:hep-ph/0101342].

\bibitem{EWPData}
The LEP Collaborations, the LEP Electroweak Working Group, 
and the SLD Heavy Flavor Group, LEPEWWG/2003-01.  \\
\url{http://lepewwg.web.cern.ch/LEPEWWG/stanmod/}

\bibitem{APV}
S.~C.~Bennett and C.~E.~Wieman,
Phys.\ Rev.\ Lett.\  {\bf 82}, 2484 (1999)
[arXiv:hep-ex/9903022].

\bibitem{Nutev}
G.~P.~Zeller {\it et al.}  [NuTeV Collaboration],
Phys.\ Rev.\ Lett.\  {\bf 88}, 091802 (2002)
[arXiv:hep-ex/0110059].

\bibitem{Heather}
The existence of the $Z_H\to W^+W^-$ decay mode was first noted by 
H.~E.~Logan,
arXiv:hep-ph/0307340.

\bibitem{bj}
J. D. Bjorken, in {\it New and Exotic Phenomena, Proceedings of the
Seventh Moriond Workshop}, O. Fackler and J. Tran Thanh Van, eds.
(\'Editions Fronti\`eres, 1987).

\bibitem{Esix}
See, e.g. J.~L.~Hewett and T.~G.~Rizzo,
Phys.\ Rept.\  {\bf 183}, 193 (1989).
We thank JoAnne Hewett and Tom Rizzo for bringing this point to our attention.


\bibitem{askIan}
Ian Hinchliffe, private communication.


\bibitem{bpdf1}
J.~C.~Collins and W.~K.~Tung,
Nucl.\ Phys.\ B {\bf 278}, 934 (1986).
\bibitem{bpdf2}
F.~I.~Olness and W.~K.~Tung,
Nucl.\ Phys.\ B {\bf 308}, 813 (1988). 
\bibitem{bpdf3}
M.~A.~Aivazis, F.~I.~Olness and W.~K.~Tung,
Phys.\ Rev.\ Lett.\  {\bf 65}, 2339 (1990).

\bibitem{Sullivan}
P.~M.~Nadolsky and Z.~Sullivan,
in {\it Proc. of the APS/DPF/DPB Summer Study on the Future 
of Particle Physics (Snowmass 2001) } ed. N.~Graf,
eConf {\bf C010630}, P510 (2001)
[arXiv:hep-ph/0110378].

\bibitem{TimSep}
D.~O.~Carlson, Ph.D. Thesis, Michigan State U., 1995;
arXiv:hep-ph/9508278.

\bibitem{TY}
T.~Tait and C.~P.~Yuan,
arXiv:hep-ph/9710372.

\end{thebibliography}
\end{document}